\documentclass[aps,pra,groupedaddress,reprint]{revtex4-1}
\usepackage[latin1]{inputenc}
\usepackage{amsmath}
\usepackage{amsfonts}
\usepackage{amssymb}
\usepackage{graphicx}

\usepackage[outercaption]{sidecap}
\usepackage{setspace}

\begin{document}
\title{Coherent Perfect Absorption: an electromagnetic perspective }
\author{Sanjeeb Dey }
\email{snjbde.1@gmail.com (Sanjeeb Dey)}
\author{Suneel Singh}
\affiliation{School of Physics, University of Hyderabad, Hyderabad-500046, India.}

\begin{abstract}
We present a simple closed expression for determining the condition for Coherent Perfect Absorption  derived through electromagnetic wave treatment and interface boundary conditions. 
Apart from providing physical insight this expression can be used to estimate the values of various parameters required for observation of coherent perfect absorption in a given medium characterized by a complex dielectric constant. The results of the theoretical expression are found to be in good agreement with those obtained through numerical simulations. 
\end{abstract}
\pacs{551564464}    
\maketitle
\section{Introduction }
The phenomenon of coherent perfect absorption\cite{cpa1,cpa2,cpa3,cpa4,cpa5,cpa6,cpa7,cpa8,cpa9,cpa10,cpa11} via symmetric illumination (at same incident angle from opposite sides) of artificially fabricated media by identical electromagnetic waves has been of tremendous interest in recent years.  As the name indicates, coherent perfect absorption ($ CPA $) is the process in which the incident coherent electromagnetic radiations are completely absorbed in a medium and appear as some other form of internal energy such as thermal or electrical energy. Such a medium is called a coherent perfect absorber or an anti lasing device as this function of the medium is exactly opposite to that of a laser  (which converts some form of incoherent energy such as thermal or electrical energy into coherent light).\\
Accordingly several schemes \cite{cpa1,cpa2,cpa4,cpa6,cpa7,cpa9,cpa10} have been proposed that involve designing of novel CPA media and investigation of $CPA$ in them. Almost all of these however, tend to be computational in nature. While computational techniques are very effective tools for investigating $CPA$ in complicated structures such as metal-dielectrics or meta-materials, they are incapable of providing insight into the nature of physical processes involved in the occurrence of CPA phenomenon. \\
For example, the essential requirement of a slightly absorbing medium (i.e., complex permittivity with positive imaginary part) for occurrence of $CPA$ is a well known fact \cite{cpa1,cpa2,cpa3,cpa4,cpa5,cpa6,cpa7,cpa8,cpa9,cpa10,cpa11} but the actual mechanism involved is not apparent. Further given a medium characterized by a dielectric constant (permittivity) $\epsilon$ it is not possible to assess or estimate the dependence of CPA on various field and medium parameters such as the incident angle, wavelength, medium composition etc.
It is thus imperative to develop a theoretical formalism for the phenomenon of $CPA$. 
Because, a theoretical expression describing dependence of the $CPA$ on various field and system parameters; apart from providing physical insight, would also be useful in determining the field parameters such as the angle of incidence, wavelength given the permittivity of the medium or vice-versa.
This article deals with such a theoretical formulation of $ CPA $ for plane electromagnetic waves. \\
The organization of the article is as follows: 
Section $II$ deals with derivation of  reflection and transmission coefficients at the interfaces of a thin slab for oblique incidences of plane electromagnetic waves using Maxwell's equations and boundary conditions on interfaces. \\ 
 Using these expressions, in section $III$ we develop the theory of CPA through derivation of an analytical expression that leads to the condition for the occurrence of $CPA$. Thereafter we  outline briefly a few schemes for observing $ CPA $ in various metal-deielctric composite media.\\
In section $IV$ we present a comparison between the  results obtained  from the analytical expression derived in previous section with those obtained through computational means for various system field parameters.
Summary and conclusions are presented in section $V$ .
\section{Geometry and formulation of CPA}
\begin{figure}[h]
\begin{center}
\includegraphics[width=6.0 cm,height=4.0 cm]{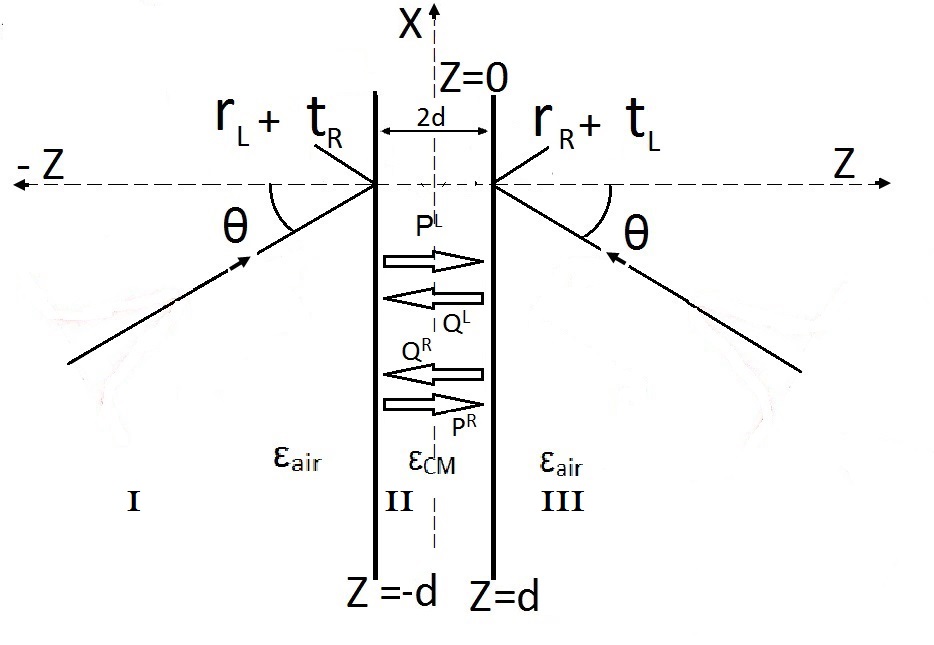} 
\caption{Two identical plane waves incident (at same angle $ \theta $) on the two  air-CPA medium interfaces \ from both sides. }
\label{fig 2}
\end{center}
\end{figure}

In a typical CPA configuration, two identical electromagnetic waves are incident from opposite direction on the left (L) and the right (R) interfaces of the CPA medium henceforth called slab. Both electromagnetic waves have the same angle of incidence $ \theta $. 
As shown in the fig. \ref{fig 2}, the two air-slab interfaces are in the $ x-y $ plane at $ z=-d $ and  $z=d $ so that the slab thickness is $ 2d $.
Owing to the symmetry of the CPA configuration and superposition principal we can consider the reflection and transmission characteristics of each electromagnetic wave separately and superpose them later to obtain the resultant transmission or reflectivity at any interface and deduce the CPA condition. 
The two $TE$ ($s$) -polarized waves are incident at oblique angle $ \theta $  upon the interfaces between air and $ CPA $ medium \cite{yeh}  at $ z=-d $ and $ z=d $. 
The incident light is partly reflected back in air, a part is transmitted through the $ CPA $ medium (to exit at the other interface) and the remaining part is reflected back and forth into the $ CPA $ medium (assuming no other losses).
The amplitude of incident, reflected and transmitted waves are denoted $ A_i$, $A_r  $ and $A_t $ respectively. This process occurs for both the incident waves.
Therefore, for incident $ TE $ plane wave on left hand side ($ LHS $) of the interface (at z=-d), the wave equations in section $I$, $II$ and $III$ shown in the Fig. \ref{fig 2} are as follows: 
\begin{eqnarray}
E_{1y}&=&(A_{in}e^{ik_{1z}z}+A_re^{-ik_{1z}z})e^{i(k_{x}x-\omega t)} 
\hspace{0.1cm} \text{for} \hspace{0.2cm} z<-d \label{eq 1}\\
E_{2y}&=&(P_{L}e^{ik_{2z}z}+Q_Le^{-ik_{2z}z})e^{i(k_{x}x-\omega t)} 
\hspace{0.1cm}\text{for}\hspace{0.1cm} -d\leq z\leq d\nonumber \\\label{eq 2}\\
E_{3y}&=&A_te^{ik_{1z}z}e^{i(k_{x}x-\omega t)} 
\hspace{0.1cm}\text{for}\hspace{0.1cm} z>d \hspace{0.1cm} \text{respectively.} \label{eq 3}
\end{eqnarray}
Here $ k_{jz}=\sqrt{k_j^2-k_{x}^2}=\sqrt{k_0^2\epsilon_j-k_{0}^2\sin^2\theta} $. Because section I and III in Fig.\ref{fig 2} are air medium, their parallel wave vector are the same, i.e. $ k_{1z} ($ for $\ j=1, 2$). 
Here $ P_L$ and $ Q_L$ are the amplitudes of the forward and backward moving waves in $ CPA $ medium (section II) as shown in Fig.\ref{fig 2}.
Similarly, one can write  $ TE $-polarized wave equations for the right hand side ($ RHS $).
\begin{figure}[h]
\begin{center}
\includegraphics[width=8.0 cm,height=5.0 cm]{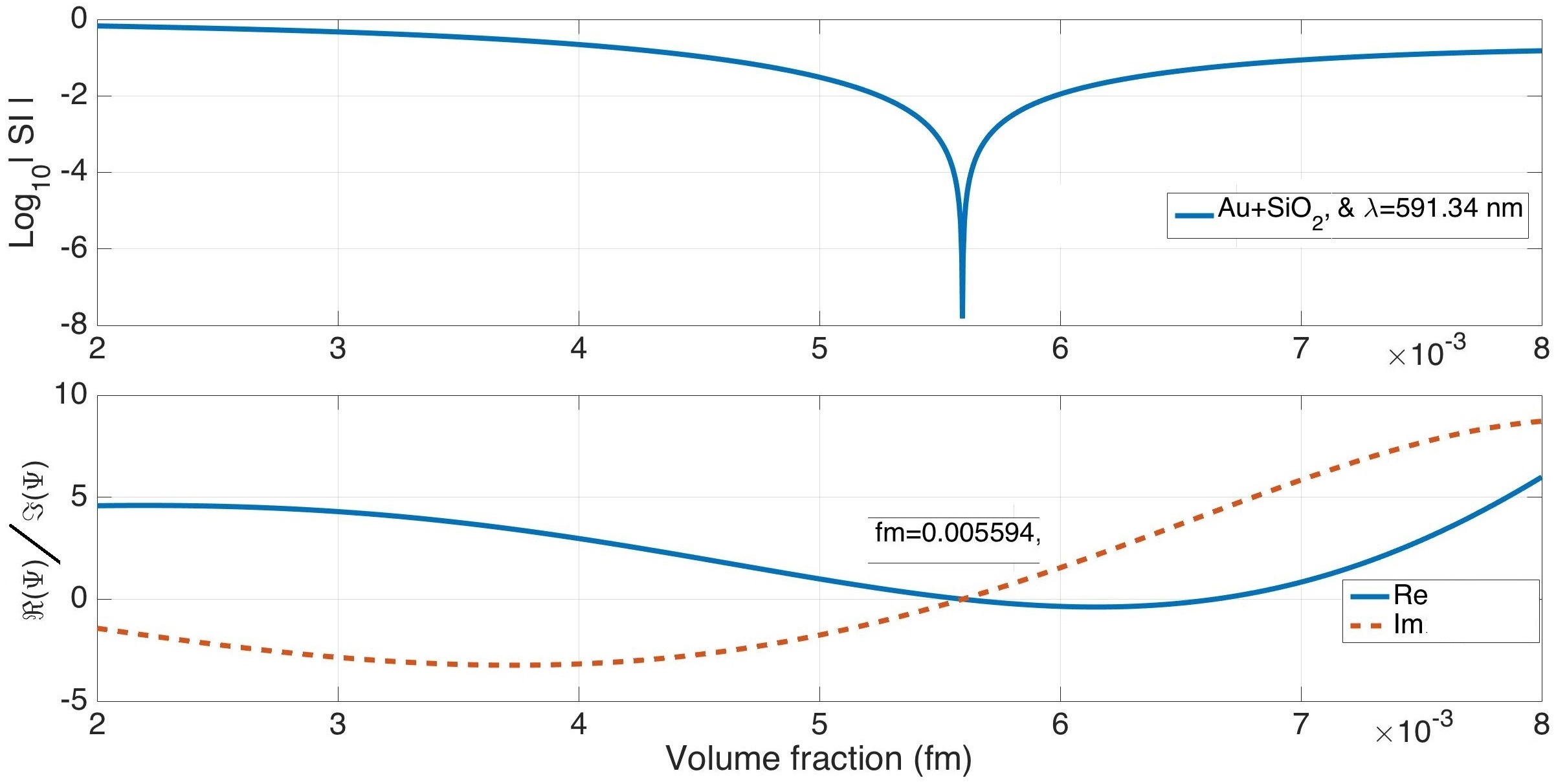} 
\caption{Top figure depicts $CPA $ as a function of volume fraction ($ f_m $) for Gold-Silica composite and the bottom figure shows that at $ CPA $ frequency, the real and imaginary parts of the function $\Psi$ become zero.  Thickness of medium is $ 2d=10 $ $ \mu m $ and angles of incidence, $ \theta=45^o $.}
\label{fig 3}
\end{center}
\end{figure}
\begin{figure}[h]
\begin{center}
\includegraphics[width=8.0 cm,height=5.0 cm]{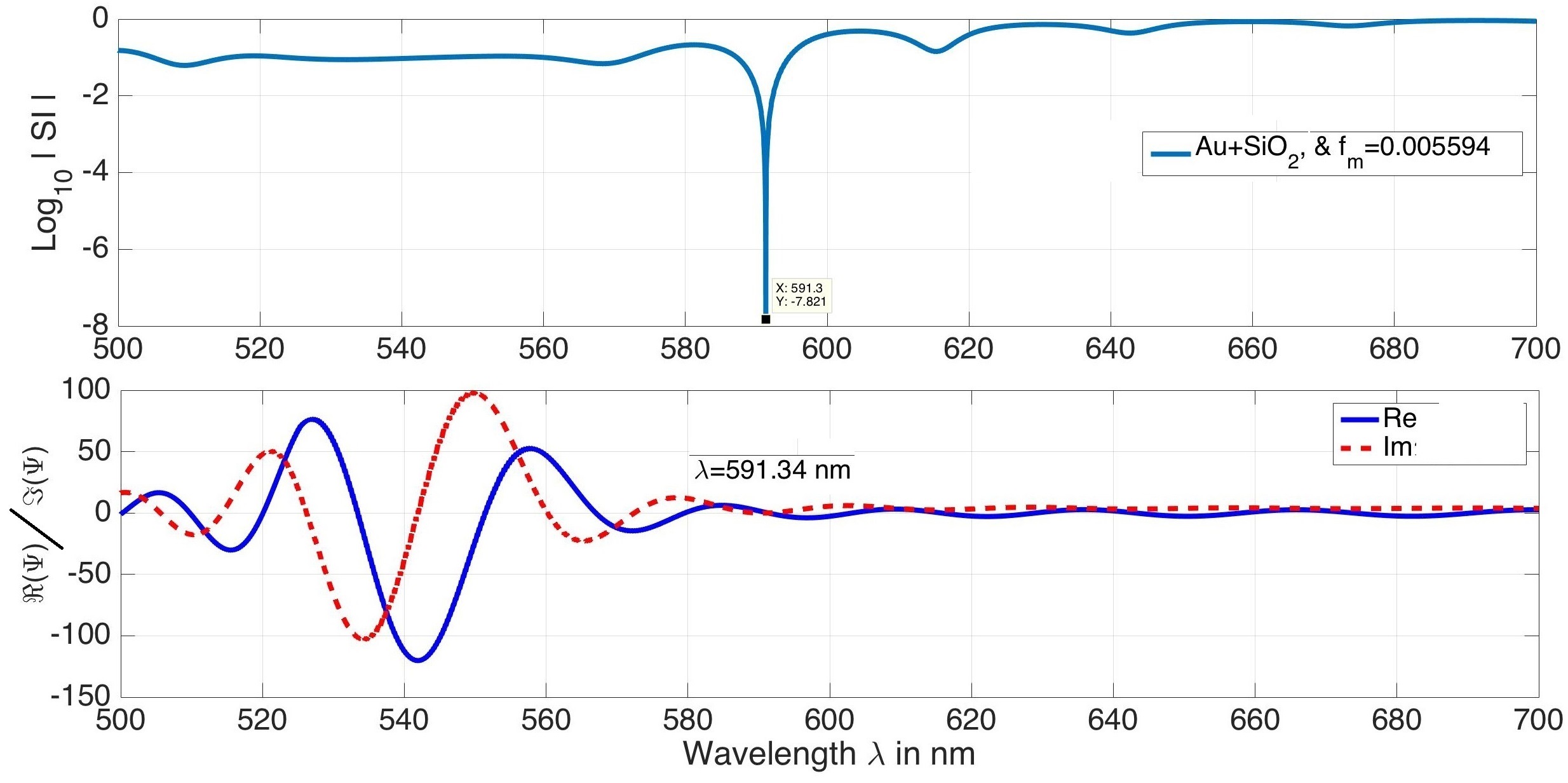} 
\caption{Top figure depicts $ CPA $ as a function of wavelength for Gold-Silica composite and the bottom figure shows that at $ CPA $ frequency, the real and imaginary parts of the function $\Psi$ become zero. Thickness of medium is $ 2d=10 $  $ \mu m$ and angles of incidence, $ \theta=45^o $.}
\label{fig 4}
\end{center}
\end{figure}
At optical frequencies the $ CPA $ medium is non-magnetic so that $ \mu=1 $ and  the corresponding magnetic field 
$\vec{H}$=($\hat{x}H_x+\hat{y}H_y+\hat{z}H_z$)$e^{i(k_{x}x-\omega t)}$
can be evaluated from the Maxwell's equation $-\frac{1}{c}\frac{\partial \vec{B}}{\partial t}=\vec{\nabla}\times\vec{E}$ as $ ik_0\vec{H}=\vec{\nabla}\times\vec{E} $. 
So, $ ik_0\vec{H}=-\hat{x}\frac{\partial E_y}{\partial z}+\hat{z}\frac{\partial E_y}{\partial x} $ where $ k_0=\frac{\omega}{c}=\frac{2\pi}{\lambda}$. Now equating the coefficients of $\hat{x}$-axis  we obtain  $ik_0H_{1x}=-\frac{\partial E_1y}{\partial z}$. So, from eq. (\ref{eq 1}), (\ref{eq 2}) and (\ref{eq 3}), one can have
\begin{figure}[h]
\begin{center}
\includegraphics[width=8.0 cm,height=4.0 cm]{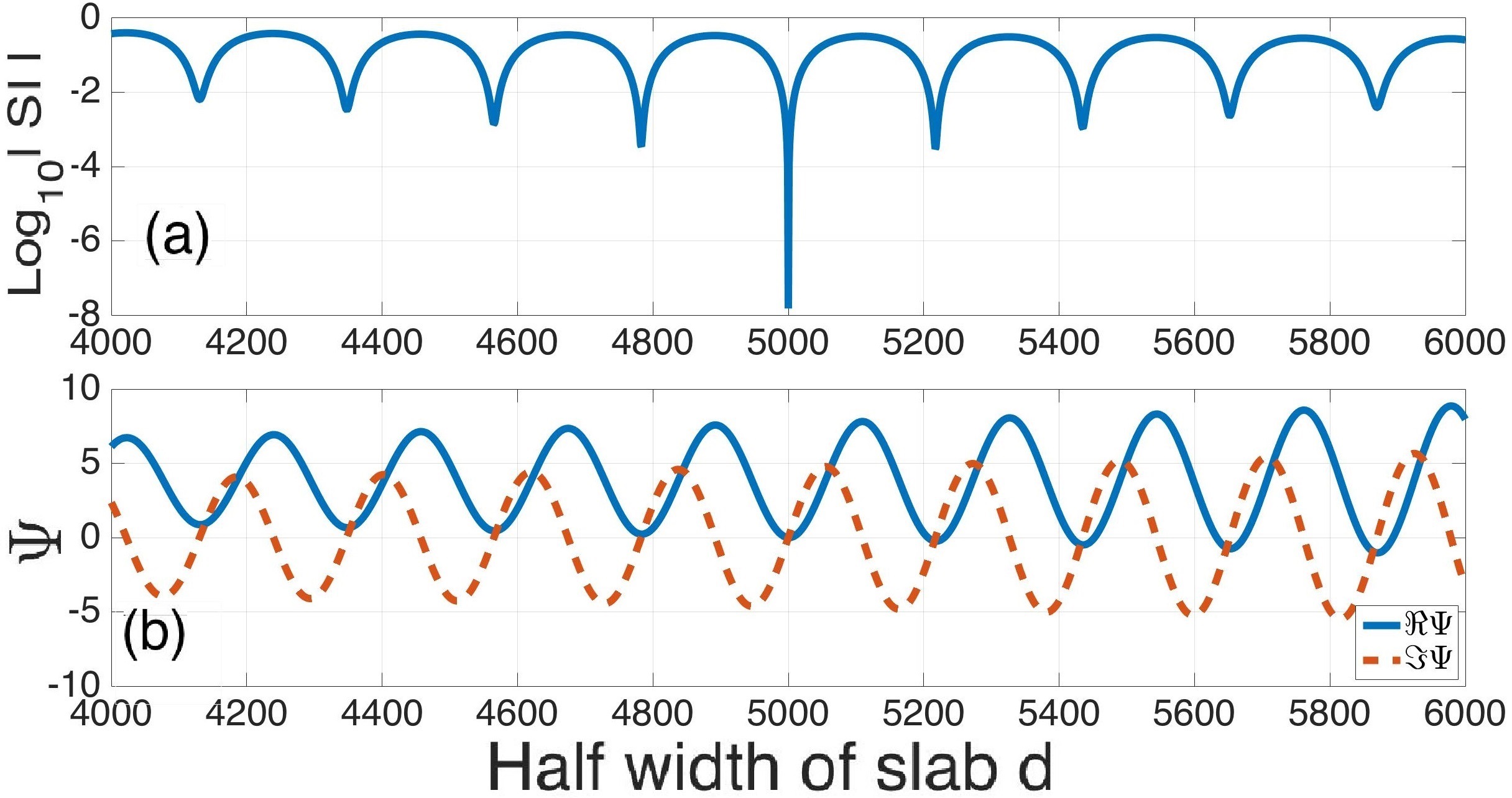} 
\includegraphics[width=8.0 cm,height=4.0 cm]{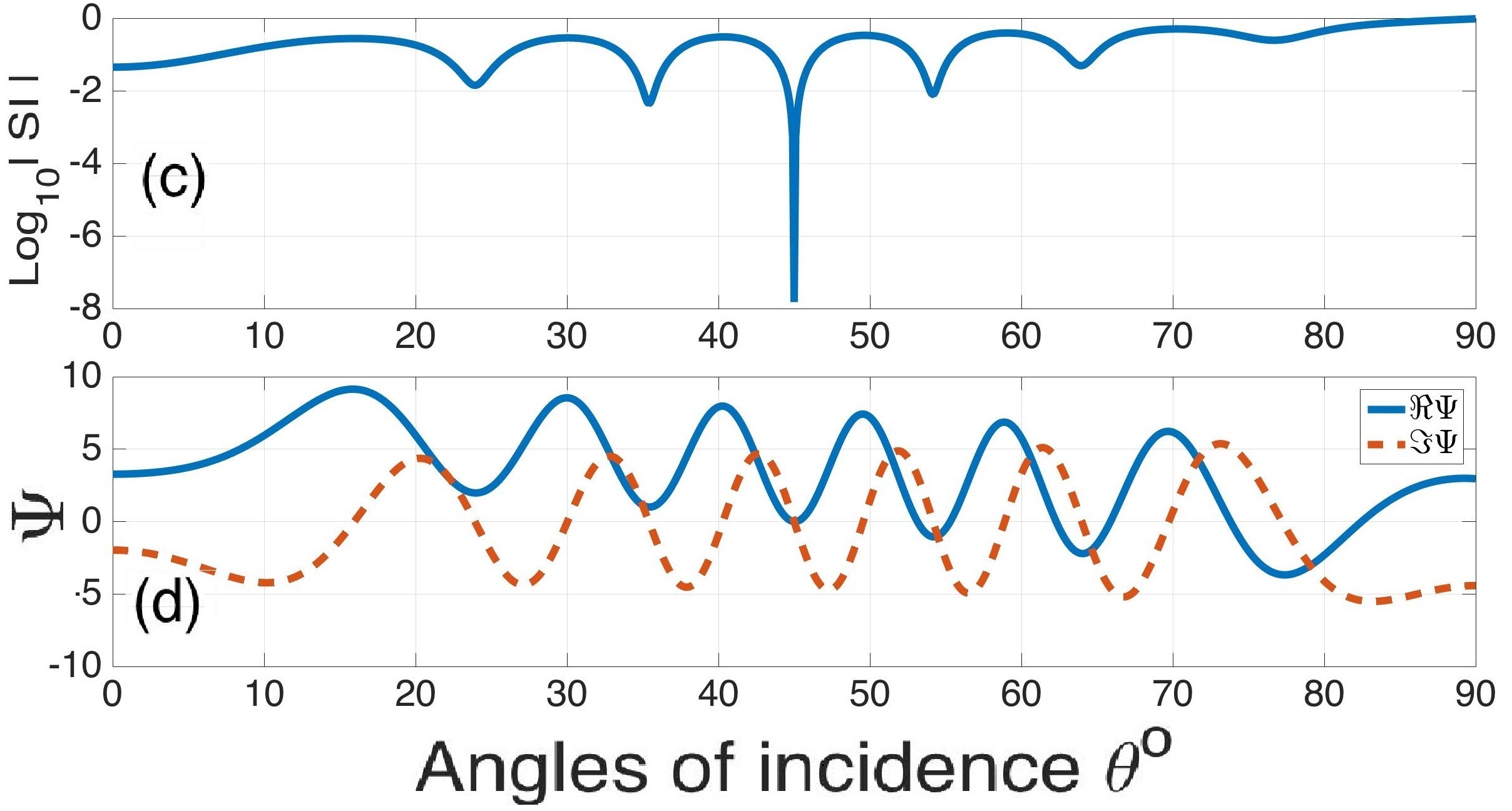} 
\caption {Top figures (a) and (c), depict $ log_{10}| SI | $ as a function of half width of the slab ($ d$ in $ nm$) and angle of incidence $ \theta^o $ respectively. 
In the bottom figures (b) and (d), $ \Re\Psi$ and $\Im\Psi $ are plotted as a function of half width of the slab ($ d$ in $ nm$) and angle of incidence $ \theta^o $ respectively.} \label{fig 4a}
\end{center}
\end{figure}
\begin{SCfigure*}
\includegraphics[width=14.0 cm,height=5.50 cm]{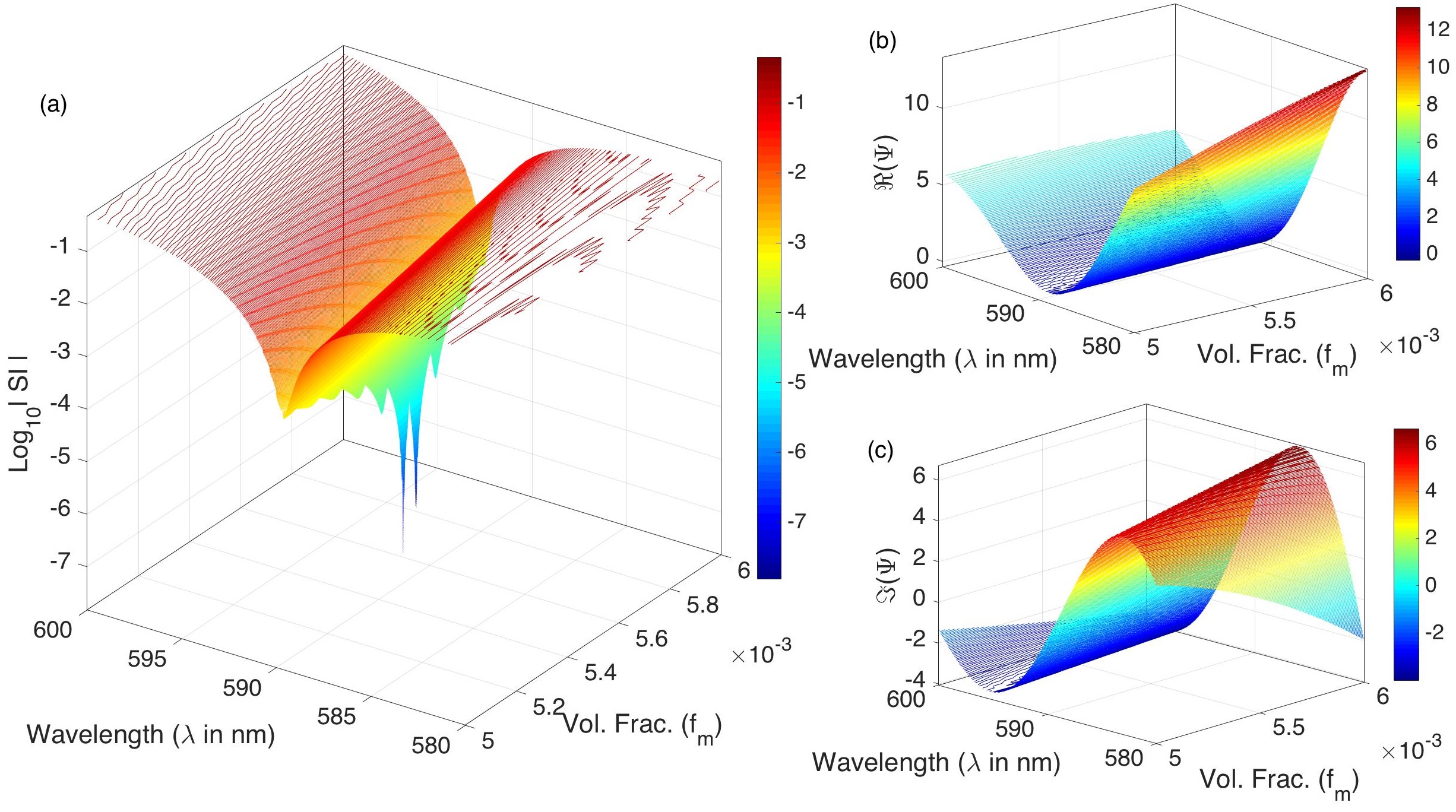} 
\caption{(a) $ Log_{10}| SI  | $ (or $ CPA $) plotted as functions of wavelength $ \lambda $  and volume fraction ($ f_m $) for Gold-Silica composite medium. (b) Real($ \Psi $) plotted as functions of wavelength $ \lambda $  and volume fraction ($ f_m $). (c) Imag($ \Psi $) plotted as functions of wavelength $ \lambda $  and volume fraction ($ f_m $). Half width of medium $ d=5 $ $ \mu m $ and angle of incidences $ \theta=45^o $.}
\label{fig 5}
\end{SCfigure*}
\begin{eqnarray}
H_{1x}&=& S_1(-A_{in}e^{ik_{1z}z}+A_re^{-ik_{1z}z})e^{i(k_{x}x-\omega t)},\hspace{0.1cm} \text{for} \hspace{0.1cm} z<-d  \label{eq 4},\\
H_{2x}&=&S_2(-P_{L}e^{ik_{2z}z}+Q_Le^{-ik_{2z}z})e^{i(k_{x}x-\omega t)},
\hspace{0.01cm}\text{for}\hspace{-0.01cm}-d\leq z\leq d,\nonumber\\\label{eq 5}\\
H_{3x}&=&S_1(-A_te^{ik_{1z}z})e^{i(k_{x}x-\omega t)}, \hspace{0.1cm} \text{for} \hspace{0.1cm} z>d, \label{eq 6}
\end{eqnarray}
where $ S_j=\frac{k_{jz}}{k_0} $ and $ j=1, 2 $.
Therefore, in section I, II and III of Fig. \ref{fig 2}, the wave equations can be written in matrix form as
\begin{eqnarray}
\begin{bmatrix}E_{1y}\\H_{1x}\end{bmatrix}&=&\begin{bmatrix}e^{ik_{1z}z} & e^{-ik_{1z}z}\\-S_1e^{ik_{1z}z}&S_1e^{-ik_{1z}z}\end{bmatrix} \begin{bmatrix} A_{in}\\A_r\end{bmatrix}  \hspace{0.1cm} \text{for} \hspace{0.1cm} z<-d \label{eq 7}  \\ 
\begin{bmatrix}E_{2y}\\H_{2x}\end{bmatrix}&=&\begin{bmatrix}e^{ik_{2z}z} & e^{-ik_{2z}z}\\ -S_2e^{ik_{2z}z} & S_2e^{-ik_{2z}z}
\end{bmatrix} \begin{bmatrix}P_L\\Q_L\end{bmatrix}   \hspace{0.1cm}\text{for}\hspace{-0.01cm}-d\geq z\leq d \nonumber\\ \label{eq 8}\\ 
 \text{and}
  \begin{bmatrix}E_{3y}\\H_{3x}\end{bmatrix}&=&\begin{bmatrix}e^{ik_{1z}z}\\-S_1e^{ik_{1z}z}\end{bmatrix} A_t  \hspace{0.1cm} \text{for} \hspace{0.1cm} z>d\hspace{0.1cm} \text{respectively.} \label{eq 9}
\end{eqnarray}
According to the electromagnetic boundary conditions at the interface $ z=d $, $ \begin{bmatrix} E_{2y}\\ H_{1x}\end{bmatrix}=\begin{bmatrix}E_{3y}\\H_{3x}\end{bmatrix} $. So,
\begin{eqnarray}
\begin{bmatrix} e^{ik_{2z}d} & e^{-ik_{2z}d}\\-S_2e^{ik_{2z}d} & S_2e^{-ik_{2z}d} \end{bmatrix}
\begin{bmatrix}P_L\\Q_L\end{bmatrix}&=\begin{bmatrix}e^{ik_{1z}d}\\-S_1e^{ik_{1z}d}\end{bmatrix}A_t \label{eq:12}
\end{eqnarray}
The  above eq.\ref{eq:12}. yields the solutions for $ P_L$ and $ Q_L $ in terms of $A_t $ as follows:
\begin{eqnarray}
 \begin{bmatrix} P_L\\Q_L\end{bmatrix} =\begin{bmatrix}(1+\frac{S_1}{S_2})\frac{e^{ik_{1z}d}}{e^{+ik_{2z}d}}\\(1-\frac{S_1}{S_2})\frac{e^{ik_{1z}d}}{e^{-ik_{2z}d}}\end{bmatrix}\frac{A_t}{2} \label{eq:13}
\end{eqnarray}
Again, the electromagnetic boundary condition at the interface $ z=-d $ yields:
\begin{eqnarray}
\begin{bmatrix}e^{-ik_{1z}d} & e^{ik_{1z}d}\\ -S_1e^{-ik_{1z}d} & S_1e^{ik_{1z}d}\end{bmatrix}\begin{bmatrix}A_{in}\\A_r\end{bmatrix}=\begin{bmatrix}e^{-ik_{2z}d} & e^{ik_{2z}d}\\-S_2e^{-ik_{2z}d} & S_2e^{ik_{2z}d}\end{bmatrix}\begin{bmatrix}P_L\\Q_L\end{bmatrix}\nonumber\\ \label{eq:15}
\end{eqnarray}
Substituting the value of $ P_L$ and $ Q_L$ from eq.(\ref{eq:13}) in the above eq. (\ref{eq:15}) we obtain
\begin{eqnarray}
\begin{bmatrix}\frac{e^{ik_{1z}d}}{2}\lbrace(1+\frac{S_1}{S_2})e^{-2ik_{2z}d}+(1-\frac{S_1}{S_2})e^{2ik_{2z}d}\rbrace & -e^{ik_{1z}d}\\\frac{e^{ik_{1z}d}}{2}\lbrace(1+\frac{S_1}{S_2})e^{-2ik_{2z}d}-(1-\frac{S_1}{S_2})e^{2ik_{2z}d}\rbrace & \frac{S_1}{S_2}e^{ik_{1z}d}\end{bmatrix}\times \nonumber \\
\begin{bmatrix}\frac{A_t}{A_{in}}\\\frac{A_r}{A_{in}}\end{bmatrix}
=\begin{bmatrix}
e^{-ik_{1z}d}\\\frac{S_1}{S_2}e^{-ik_{1z}d}\end{bmatrix} \hspace{01.5cm} \label{eq:16} 
\end{eqnarray}
Here, $ t_L=\frac{A_t}{A_{in}} $ and $r_L= \frac{A_r}{A_{in}} $ respectively, are the transmission ($ t_L $) and reflection ($ r_L $) coefficient that can be obtained from the solution of the following matrix equation
\begin{eqnarray}
&\begin{bmatrix}t_L\\r_L\end{bmatrix}=D^{-1}\begin{bmatrix}
e^{-ik_{1z}d}\\\frac{S_1}{S_2}e^{-ik_{1z}d}\end{bmatrix} \text{, } \quad \text{where} \qquad\label{eq:17} \\
 &D=\begin{bmatrix}\frac{e^{ik_{1z}d}}{2}\lbrace(1+\frac{S_1}{S_2})e^{-2ik_{2z}d}+(1-\frac{S_1}{S_2})e^{2ik_{2z}d}\rbrace & -e^{ik_{1z}d}\\\frac{e^{ik_{1z}d}}{2}\lbrace(1+\frac{S_1}{S_2})e^{-2ik_{2z}d}-(1-\frac{S_1}{S_2})e^{2ik_{2z}d}\rbrace & \frac{S_1}{S_2}e^{ik_{1z}d}\end{bmatrix} \nonumber 
 \end{eqnarray}
Using $ D^{-1} $  and from  eq.s (\ref{eq:17}) the solution for $ t_L$ and $r_L $ are 
\begin{eqnarray}
t_L&=&\frac{4\frac{S_1}{S_2}e^{-2ik_{1z}d}}{\lbrace(1+\frac{S_1}{S_2})^2 e^{-2ik_{2z}d}-(1-\frac{S_1}{S_2})^2e^{2ik_{2z}d}\rbrace} ,\quad\label{eq23}\\
r_L&=&\frac{e^{-2ik_{1z}d}(1-\frac{S_1^2}{S_2^2})(e^{2ik_{2z}d}-e^{-2ik_{2z}d})}{\lbrace(1+\frac{S_1}{S_2})^2 e^{-2ik_{2z}d}-(1-\frac{S_1}{S_2})^2e^{2ik_{2z}d}\rbrace}, \label{eq24}
\end{eqnarray} 
respectively. Similarly one can obtain the solutions for an electromagnetic wave incident from $RHS$ on interface at $ z=d $ and these are found to be similar as $ t_R= t_L$ and $ r_R = r_L$ shown in Eq.(\ref{eq23}) and Eq.(\ref{eq24}).
Now we have all reflection ($r_L $ and $r_R $) and transmission ($ t_L$ and $t_R $) coefficients for  waves incident on both left and right side interfaces of the slab.
In the next section we determine the condition for $CPA$.
\begin{SCfigure*}[]
\includegraphics[width=14.0 cm,height=5.0 cm]{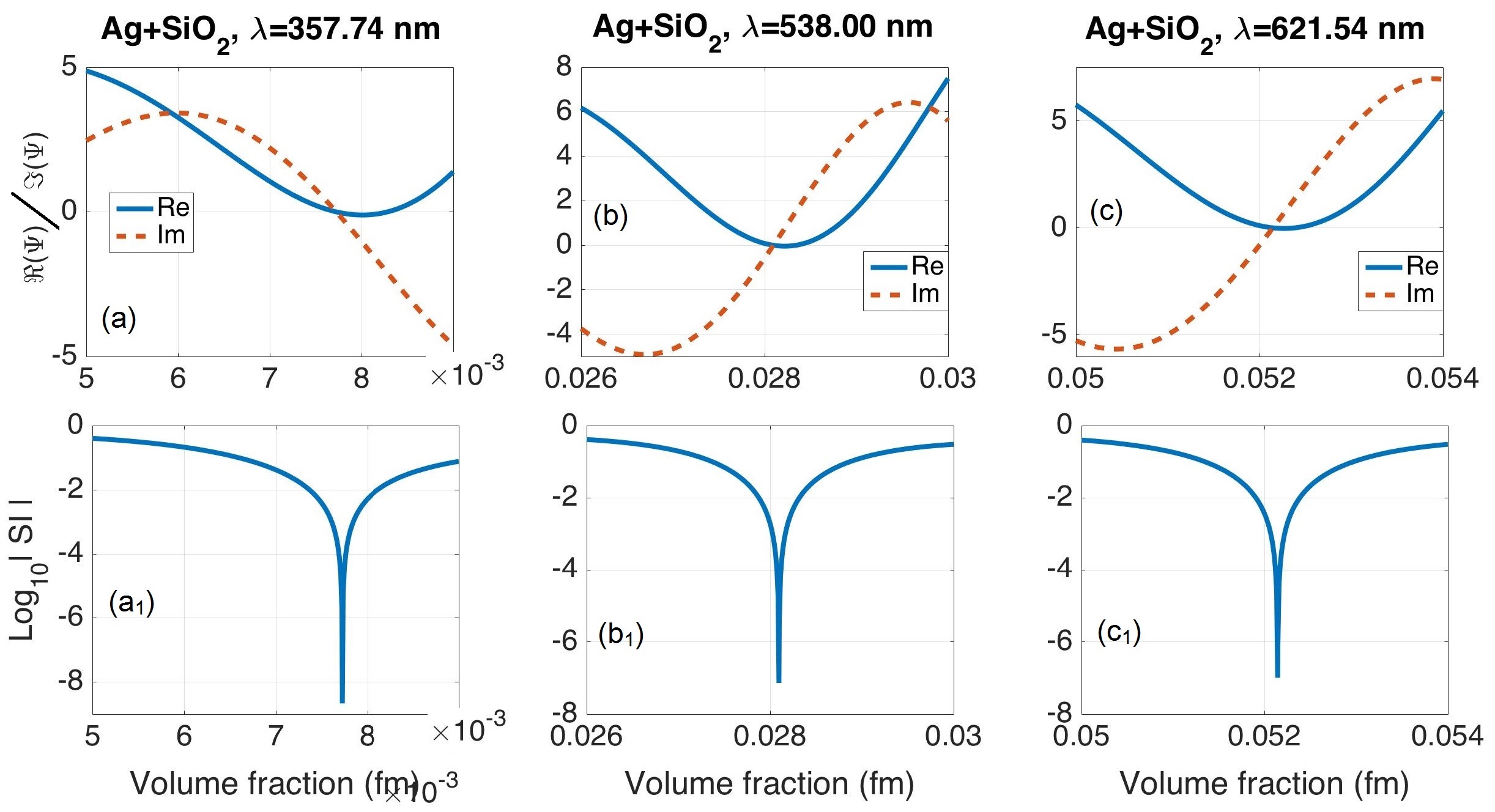} 
\caption{Top row shows $ CPA $ for different frequencies as a function of volume fraction ($ f_m $) for Silver-Silica composite and the bottom row shows that at $ CPA $ frequency, the real and imaginary parts of the function $\Psi$ become zero. Here, half thickness of the medium is, $ d=5 \mu m $ and angle of incidences, $ \theta_i=45^o $. }
\label{fig 6}
\end{SCfigure*}
\section{ CPA Condition}
As shown in Fig.\ref{fig 2} the field exiting the medium at either interface has contributions from both forward (wave on LHS) and the backward (wave on RHS). Thus for CPA to occur, the coefficient of the field at the interface z = -d (or z = d) given by $r_L+t_R $  (or $r_R+t_L$) must vanish, that is:\\
$ \frac{e^{-2ik_{1z}d}(1-\frac{S_1^2}{S_2^2})(e^{2ik_{2z}d}-e^{-2ik_{2z}d})}{\lbrace(1+\frac{S_1}{S_2})^2 e^{-2ik_{2z}d}-(1-\frac{S_1}{S_2})^2e^{2ik_{2z}d}\rbrace} + $
\begin{eqnarray} 
\frac{4\frac{S_1}{S_2}e^{-2ik_{1z}d}}{\lbrace(1+\frac{S_1}{S_2})^2 e^{-2ik_{2z}d}-(1-\frac{S_1}{S_2})^2e^{2ik_{2z}d}\rbrace}=0 \label{eq:20}
\end{eqnarray}
Using the relations $ S_1=k_{1z}/k_0$, $ S_2=k_{2z}/k_0$ and $ e^{2ik_{2z}d}-e^{-2ik_{2z}d}=2i\sin(2k_{2z}d) $ in above Eq.(\ref{eq:20}) and upon simplification, we arrive at the final expression governing the occurrence of CPA in the medium II
\begin{eqnarray}
2k_{1z}k_{2z}+i(k_{2z}^2-k_{1z}^2)sin(2k_{2z}d)=0 .\label{eq 20a}
\end{eqnarray}

The above expression Eq.(\ref{eq 20a}) is a function (through $k_{1z}$, $k_{2z}$), of the permittivities $\epsilon_j$, j=1, 2, 3 of medium I, II and III (medium I and III are air in the present case), angle of incidence $\theta$ and wavelength  $\lambda$ of the incident waves and slab thickness, 2d.  Thus given  the above parameters, it is possible to determine the permittivity of the medium II under CPA condition. Here it should also be noted  that though all the given parameters are real quantities in the present case, the solution of above Eq.(\ref{eq 20a}) for propagation constant $k_{2z}$ ( or $\sqrt{\epsilon_2}$) in the medium will be complex. In other words, the expression Eq.(\ref{eq 20a}) demonstrates that an essential condition for  CPA to occur is: the medium must definitely be dissipative. 
Thus if we write the expression Eq.(\ref{eq 20a}) as a complex function $\Psi=2k_{1z}k_{2z}+i(k_{2z}^2-k_{1z}^2)sin(2k_{2z}d)$ it is observed that in order to fulfil the CPA condition stipulated in Eq.(\ref{eq 20a}), both ($ \Re $) and imaginary ($ \Im $) part of $ \Psi $  have to be simultaneously zero for given  $ d $, $ \theta $, $ \lambda$ and $\epsilon_j$ (j=1, 2, 3). \\.
\textbf{$ CPA $ medium:} For verification of theoretical expression, we now choose metal-dielectric composite medium ($ CM $) which is a homogeneous mixture of Gold ($ Au $) or Silver ($ Ag $) with Silica ($ SiO_2 $). To calculate $ CM $ permittivity we use the Bruggeman effective medium theory \cite{cpa6,cpa9,cm,cm1,cm2}($ BEMT $), which is as follows \\$ \epsilon_{CM}=\frac{1}{4}\lbrace(3f_1-1)\epsilon_1+(3f_2-1)\epsilon_2  $
\begin{eqnarray}
\pm  \sqrt{[(3f_1-1)\epsilon_1+(3f_2-1)\epsilon_2]^2+8\epsilon_1\epsilon_2}\rbrace \label{eq:22}
\end{eqnarray}
Here, $ \epsilon_1 $, ($ f_1 $) and $ \epsilon_2 $, ($ f_2=1-f_1 $) are permittivity (filling factor) of metal and dielectric respectively.
Experimental parameters of Jhonson and Christy  for $ Au $ or $ Ag$ \cite{met}  are used in Eq.(\ref{eq:22}) to estimate permittivity of metal-dielectric composite (Gold composite, $ GC=Au+SiO_2 $ or Silver composite, $SC= Ag+SiO_2 $)  
which is useful for calculation of  $ CPA $  medium (as given in refs. \cite{cpa6,cpa9,cpa10}).
Here, width of $ CPA $ medium is $ 2d=10 $ $ \mu m $ and angles of incidence, $ \theta=45^o $  are fixed for all $ CPA $ examples.
The various scattering intensities ($ SI $) given by square of scattering amplitudes ($ SA $) are denoted by $ SA_{LHS}=r_L+t_R $ at $ z=-d$ and  $ SA_{RHS}=r_R+t_L $ at $z=d $ in Fig. \ref{fig 2}. These are numerically calculated from Fresnel's formulas of reflection and transmission coefficients.
As is well known, at $ CPA $ frequency the $ SI $ reduces to zero. 
To determine it's magnitude and freuency at which $ CPA $ occurs, we calculate $ Log_{10} |SI| $ \cite{cpa1,cpa2,cpa3,cpa4,cpa5,cpa6,cpa7,cpa8,cpa9,cpa10,cpa11}.
Now it is time to verify the results obtained from theoretical expression using (function $\Psi$) with those obtained through numerical simulations of $ CPA $.
Under the condition of $ CPA $ the real part ($\Re(\Psi$)) and the imaginary part ($ \Im(\Psi) $) of the function $ \Psi $ given by Eq.(\ref{eq 20a}) should simultaneously reduce to zero.
\begin{SCfigure*}[]
\includegraphics[width=14.0 cm,height=5.0 cm]{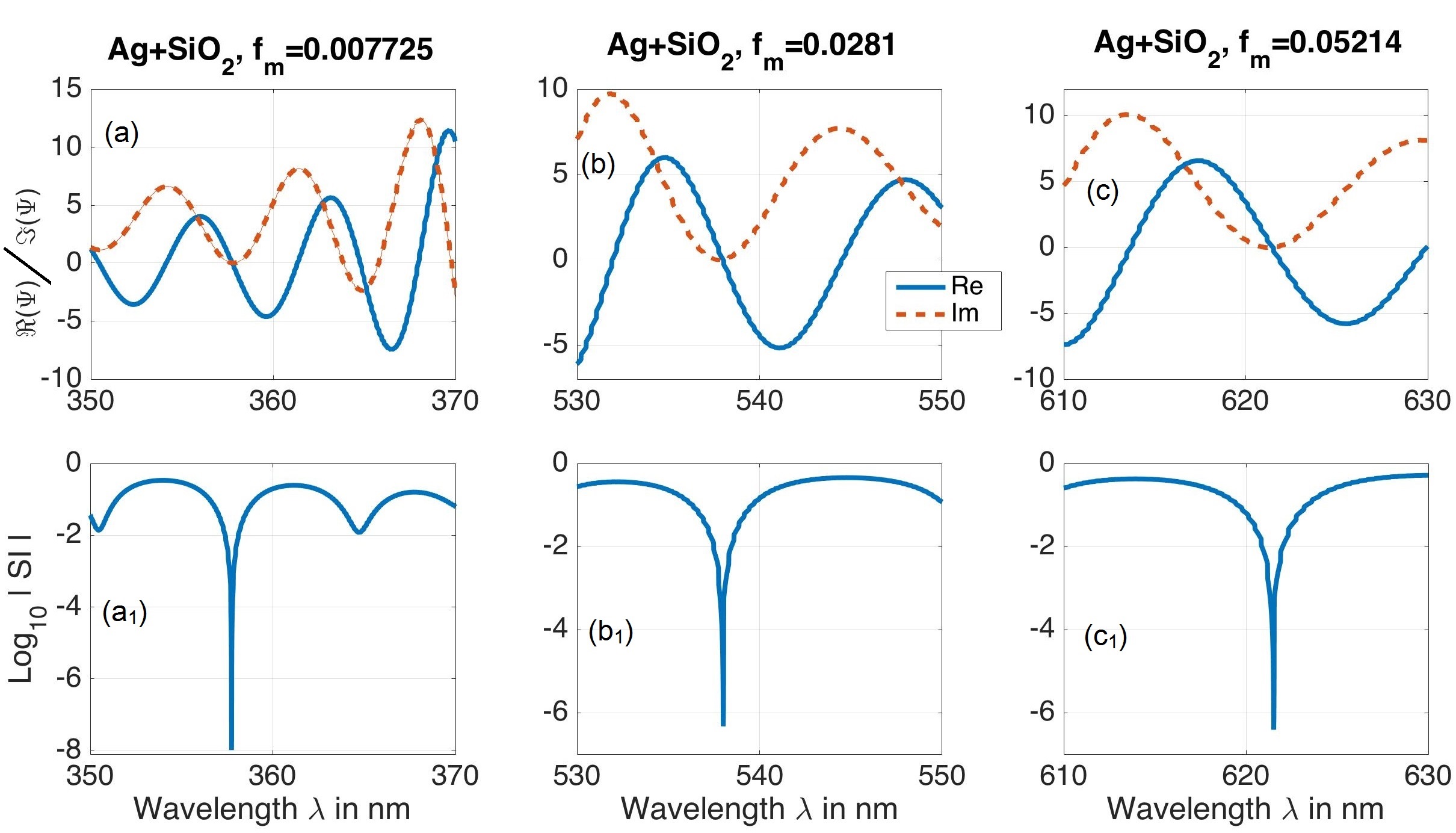} 
\caption{Top row shows $ CPA $ for different volume fraction as a function of wavelengths for Silver-Silica composite and the bottom row shows that at $ CPA $ frequency, real and imaginary part of the function $\Psi$ become zero. Here, thickness of the medium $ 2d=10 $  $\mu m $ and angle of incidences $ \theta_i=45^o $.  }
\label{fig 7}
\end{SCfigure*}
\section{Results and discussion}
We now verify the results of function $ \Psi $ (eq. \ref{eq 20a}) with $CPA$ results obtained through numerical computation and depicted in fig. \ref{fig 3} and \ref{fig 4}.

In fig. \ref{fig 3}(a) $log_{10}| SI |$ is plotted as function of volume fractions ($ f_m $) and incident wavelength ($ \lambda $) is $ 591.34 $ $ nm $; which gives $ CPA $ at (metal) volume fraction $ f_m=0.005594 $ .
Now in fig. \ref{fig 3}(b), $ \Re\Psi $ and $ \Im\Psi $ are plotted as a function of volume fraction and it is observed that at the wavelenth where $ CPA $ occurs, both are simultaneously zero as they cross the x axis.
In fig. \ref{fig 4}(a) at $ f_m=0.005594 $ on $ GC $, $log_{10}| SI |$ is plotted as a function of wavelength ($ \lambda $) and $ CPA $ is observed at $ \lambda=591.34 $ $ nm $.
In fig. \ref{fig 4}(b) $ \Re\Psi $ and $ \Im\Psi $ are plotted as a function of wavelength and it is found  they  simultaneously cross zero line exactly at $ 591.34 $ $ nm $.
We can see characteristics of $ \Re\Psi $ and $ \Im\Psi $ for $ GC $ while both volume fraction and wavelength simultaneously vary in the fig. \ref{fig 5}(b) and \ref{fig 5}(c) respectively where as fig. \ref{fig 5}(a) shows $ log_{10}|SI| $ under the same condition.\\
It is observed that at every $ CPA $ point $ \Re\Psi $ and $ \Im\Psi $ goes to zero simultaneously. An obvious question that arises is that if at some point $ \Re\Psi $ and $ \Im\Psi $  value is zero simultaneously for any arbitrary element then can we claim $ CPA $ at that point? \\
So, we choose $ SC $ a different composite medium for this test.
We tested three different wavelengths ($ 357.74$ $nm $, $538.00 $  $nm $ and $621.54 $ $nm $) as shown in the fig. \ref{fig 6}(a), fig. \ref{fig 6}(b) and fig. \ref{fig 6}(c) respectively which are depicted $ \Re\Psi $ and $ \Im\Psi $ as a function of volume fraction of $ SC $.
It is found that at volume fraction ($ f_m =$) $0.007725 $, $ 0.028100 $ and $0.052140 $ respectively they are crossing the zero line simultaneously.
And we have found, exactly at same volume fractions $ log_{10}|SI| $ shows $ CPA $ for these incident wavelengths as shown in the fig. \ref{fig 6}($ a_1 $), fig. \ref{fig 6}($ b_1 $) and fig. \ref{fig 6}($c_1$) respectively.
Now in the fig. \ref{fig 7}($ a $), fig. \ref{fig 7}($ b $) and fig. \ref{fig 7}($ c $) the $ f_m $ are fixed at $0.007725 $, $ 0.028100 $ and $0.052140 $ for $ SC $ and observed $ \Re\Psi $ and $ \Im\Psi $ as a function of wavelength, which  at $ 357.74$ $nm $, $538.00 $  $nm $ and $621.54 $ $nm $ respective wavelengths are crossing the x- axis (zero line).
So, we plot $ Log_{10}|SI| $ as a function of wavelengths for same $ f_m $ values of $ SC $ in the fig. \ref{fig 7}($ a_1 $), fig. \ref{fig 7}($ b_1 $) and fig. \ref{fig 7}($ c_1 $) respectively.
And we observe $ CPA $ exactly at the same three different wavelengths.
\section{Conclusion}
In this article we have developed a theoretical formulation of $CPA$ in order to gain insight into the complicated process of $ CPA $ and to explain the numerical results of $CPA$ in composite mixtures of metals and dielectric. 
Excellent agreement between the results obtained from the theoretical expression and the numerically computed results of $CPA$ for various composite medium is observed. 
From theoretical expression, it is clear that at $ CPA $ the real ($ \Re\Psi $) and imaginary ($ \Im\Psi $) part would always simultaneously reduce to zero.
Therefore, on the other hand one can remark that $ CPA $ could be achieve when $ \Re\Psi=0 $ and $ \Im\Psi=0 $ simultaneously.
The function $ \Psi $ is very useful for both theoretical and experimental studies, because it gives all parameter details required for observation of $ CPA $ phenomenon in a medium.
	
\bibliography{basename of .bib file}

%
%
\end{document}